\documentclass[preprintnumbers]{revtex4}
\UseRawInputEncoding
\usepackage{amssymb}
\usepackage{amsmath}
\usepackage{graphicx}
\usepackage{dcolumn}
\usepackage{bm}
\usepackage{subfigure}
\usepackage{color}

\begin{document}

\title{Shadow thermodynamics of non-linear charged Anti-de Sitter black holes}
\author{Yun-Zhi Du$^1$, Huai-Fan Li$^1$\footnote{the corresponding author}, Xiang-Nan Zhou$^2$, Wei-Qi Guo$^2$, Ren Zhao$^1$}

\address{$^1$Institute of Theoretical Physics, Shanxi Datong University, Datong,  China\\
$^2$ College of Physics and Information Engineering, Shanxi Normal University, TaiYuan, China}

\thanks{\emph{e-mail:duyzh22@sxdtdx.edu.cn, huaifan999@163.com,zhouxn10@let.edu.cn,guo970527@163.com,zhao2969@163.cn}}

\begin{abstract}
Non-linear interaction between the electromagnetic fields (EMF) are occurred when vacuum polarization in quantum electrodynamics (QED) happens. The field of non-linear electrodynamics which may be resulting from this interaction could have important effects on black hole physics. In this work, we investigate the relationship between the observable quantity, the shadow radius and the first-order phase transition for the non-linear charged AdS black hole in the frame of the Einstein-power-Yang-Mills (EPYM) gravity. Through the analysis, we find with a certain condition there exist the non-monotonic behaviors between the shadow radius, the horizon radius, and temperature (or pressure). And from the viewpoint of the shadow radius, the phase transition temperature is higher than that from the viewpoint of the horizon radius with the same condition. These indicate that the shadow radius can be regarded a probe to reveal the thermodynamic phase transition information of black holes. When the system is undergoing the phase transition in two cases of the different non-linear YM charge parameter values: $\gamma=1,~1.5$, the thermal profiles of the coexistent big and small black hole phases with the temperature are presented. Furthermore, the effects of non-linear YM charge parameter on the shadow radius and the thermal profile are also investigated.
\end{abstract}

\maketitle

\section{Introduction}

In universe there exists a most interesting and bizarre object, black hole. After a lot of researches in the past years, people believe in the existence of supermassive black holes in heart of the most galaxies. At least we can address to the two of them as the Milky Way and neighboring elliptical $M87^*$ galaxies \cite{Gillessen2017,Avery2015}. The breakthroughs on the existence of black hole are the Event Horizon Telescope (EHT) Collaboration who announced the first and the second images of the supermassive black holes $M87^*$ \cite{Akiyama2019,Akiyama2019a,Akiyama2019b} and the supermassive black holes $SgrA^*$ \cite{Akiyama2022,Akiyama2022a}. For this purpose, they used a technique called Very Long Baseline Interferometry (VLBI). As well as we know, the gravitational attraction around a black hole is so intense that the nearby objects will fall into it when they reach at a critical radius. These astounding phenomena are known as the gravitational lensing. It is expected that we can understand the properties of black holes more better from the lensing effect. That leads to the more considerable attention focused on the gravitational lensing in the strong gravity regime of black holes. Some lensing observable of Kerr black hole have been evaluated in \cite{Bozza2002,Vazquez2004,Bozza2004} and more discussions can be found in \cite{Beckwith2005,Wei2012,Chen2010,Gralla2020,Hsiao2020,Islam2021} and therein.

In addition, when the nearby objects are photons emitted by an illuminated source located behind the black hole, then it brings a shadow which can be seen by an observer sitting at infinity. The shadow concept has been studied firstly by Bardeen \cite{Bardeen1973}. Specifically, the photons that escape from the spherical orbits form the boundary of the dark silhouette of the black hole. This dark silhouette is known as black hole shadow from the outside observer. Note that the shadows of spherically symmetric black holes are circular \cite{Synge1966,Chandrasekhar1998,Eiroa2019,Belhaj2020,Belhaj2022}, but for the spinning black holes their shadows are actually deformed \cite{Chandrasekhar1992,Falcke2000,Belhaj2021,Belhaj2021a,Belhaj2021b,Belhaj2021c}. Moreover, the photon region of gravitational lensing also provides key properties in the black hole shadow. And the unstable photon regions outside the black hole event horizon can realize the possibility to observe the black hole directly. The first black hole image of $M87^*$ published by EHT collaborations in 2019 had given the certain constraints on some shadow observable. Later the angular shadow radius of $M87^*$ was addressed in \cite{Kocherlakota2021}. Recently the angular shadow radius of $SgrA^*$ in the second black hole image has been presented \cite{Akiyama2022,Akiyama2022a}. It is noted that these constraints on the shadow radius are obtained with the Kerr geometry in GR as a premise, but they cannot exclude other black holes in GR or some exotic black holes in the modified gravity theory. The EHT observation of shadow can be applied as a tool to constrain the parameters of various black holes in various theories of gravity, and in the future more precise observational results can even be useful to distinguish different black holes. Therefore, we should show the theoretically possible results of various black holes to be tested by the future observational data of EHT.

On the other hand, the black hole thermodynamics has attracted more attention because of the remarkable and non-trivial results corresponding to the criticality and the stability behaviors. A crucial issue has been devoted to the investigation of the black holes in AdS/dS geometry \cite{Kubiznak2012,Altamirano2013,Kubiznak2016,Marks2021,Cai2013,Wei2013,Banerjee2011,Hendi2017,Bhattacharya2017,Wei2021}. The needed quantities have been nicely computed for various gravity theories. Considering the cosmological as a pressure, certain black holes show similarities with van der Waals fluid systems where some universalities have been obtained. Furthermore, Hawking and Page had shown the theoretical evidence of the certain transition in the phase space for the Schwarzschild-AdS black hole \cite{Hawking1983}.  The first-order phase transition of the charged Reissner Nordstrom-AdS (RN-AdS) black hole has been studied in different backgrounds including Dark Energy (DE) and Dark Matter (DM) \cite{Belhaj2022,Belhaj2020a,Belhaj2019}. In this way, the shadows and the deflection angle of the light rays by black holes in arbitrary dimensions have been discussed in \cite{Belhaj2021,Belhaj2022,Belhaj2020a,He2022}. The effect of DE and the space-time dimension on the involved optical quantities has been inspected. In Refs. \cite{Belhaj2021a,Belhaj2021c}, for instance, the influence of DM on the shadows and the photon rings of a stringy black hole illuminated by certain accretions has been also studied. Motivated by the investigations on these spacetimes, interplays between the black hole thermodynamics and the optical properties have been established. It has been shown that the shadow size can provide information on the HP phase transitions, the critical behaviors and the microstructure states of the black holes living in AdS geometries \cite{Zhang2020,Belhaj2020b,Cai2021}. And the relation between the shadow and the thermodynamics of the black hole has been also developed for regular space-times \cite{Guo2022}. Using the elliptic function analysis, it has been explored further a fundamental connection between the AdS black hole thermodynamics and the deflection angle of the light rays. Concretely, various thermodynamics behaviors of such black holes have been approached in terms the deflection angle variations \cite{BelhajPLB}. However, there is little work about the concrete relation between two coexistent black hole phases and their shadows for a black hole undergoing the first-order phase transition. In this work, we will investigate this issue for the non-linear charged AdS black hole.

The linear charged black holes in AdS spacetime \cite{Chamblin1999} within a second-order phase transition show a scaling symmetry: at the critical point the state parameters scale respects to charge q, i.e., $S\sim q^2,~P\sim q^{-2},~T\sim q^{-1}$ \cite{Johnson2018}. It is naturally to gauss whether there exists the scaling symmetry in the non-linear charged AdS black holes. As a generalization of the charged AdS Einstein-Maxwell black holes, it is interesting to explore new non-linear charged systems. Due to infinite self-energy of point like charges in Maxwell's theory \cite{Born1934,Kats2007,Anninos2009,Cai2008,Seiberg1999}, Born and Infeld proposed a generalization when the field is strong, bringing in non-linearities \cite{Dirac2013,Birula1970}.
The power-law Maxwell electrodynamics is a famous of nonlinear electrodynamics models, which involves a Yang-Mill field exponentially coupled to Einstein gravity. As well as we know, the electromagnetic part of its Lagrangian density is $Tr[F_{\mu\nu}^{(a)}F^{\mu\nu(a)}]^\gamma$. There possesses the conformal invariance and is easy to construct the analogues of the four-dimensional Reissner-Nordstr\"{o}m black hole solutions in higher dimensions. Additionally several features of the Einstein-power-Yang-Mills (EPYM) gravity in extended thermodynamics have recently been studied \cite{Zhang2015,Moumni2018,Du2021}.

Inspired by these, we mainly investigate the relation between shadow radius and phase transitions of the charged EPYM AdS black hole. This work is organized as follows. In Sec. \ref{scheme2}, we would like to briefly review the thermodynamic quantities and present the relationship between the observable quantity (i.e., shadow radius) and the horizon radius of the charged EPYM AdS black hole. In Sec. \ref{scheme3}, the behaviour of the shadow radius with the horizon radius and temperature nearby the phase transition point are investigated. And the effect of non-linear YM charge parameter on these behaviors is also presented. In order to exhibit more intuitively the relationship between the EPYM AdS black hole phase structure and its shadow, we show the thermal profiles of two coexistent big and small black hole phases in the two-dimensional plane, see in Sec. \ref{scheme4}. A brief summary is given in Sec. \ref{scheme5}.

\section{Shadow of Non-Linear charged AdS black holes}
\label{scheme2}

The solution of the four-dimensional Einstein-power-Yang-Mills (EPYM) AdS black hole with non-linear Yang-Mills (YM) charge was presented in Ref. \cite{Yerra2018}:
\begin{eqnarray}
d s^{2}=-f(r) d t^{2}+f^{-1} d r^{2}+r^{2} d \Omega_{2}^{2},\\
f(r)=1-\frac{2 M}{r}-\frac{\Lambda}{3} r^{2}+\frac{\left(2 q^{2}\right)^{\gamma}}{2(4 \gamma-3) r^{4 \gamma-2}},\label{fr}
\end{eqnarray}
where $d\Omega_{2}^{2}$ is the metric on unit $2$-sphere with volume $4\pi$ and $q$ is the YM charge. Note that this solution is valid for the condition of the non-linear YM charge parameter $\gamma\neq0.75$, and the power YM term holds the weak energy condition (WEC) for $\gamma>0$ \cite{Corda2011}. In the extended phase space, $\Lambda$ was interpreted as the thermodynamic pressure $P=-\frac{\Lambda}{8\pi}$. The black hole event horizon locates at $f(r_+)=0$.

The thermodynamical quantities of this system are
\begin{eqnarray}
M&=&\frac{1}{6}\left[8 \pi Pr_+^{3 / 2}+3r_+^{\frac{3-4 \gamma}{2}} \frac{\left(2 q^{2}\right)^{\gamma}}{8 \gamma-6}+3 r_+\right].\label{M}\\
T&=&\frac{1}{4 \pi r_{+}}\left(1+8 \pi P r_{+}^{2}-\frac{\left(2 q^{2}\right)^{\gamma}}{2 r_{+}^{(4 \gamma-2)}}\right),~~~~~S=\pi r_{+}^{2},~~~~V=\frac{4\pi r_+^3}{3}.    \label{T}\\
P&=&r_+\left[\frac{T}{2}-\frac{1}{8 \pi}r_++\frac{\left(2 q^{2}\right)^{\gamma}}{16 \pi}r_+^{1-4\gamma}\right].\label{PTV}
\end{eqnarray}
It is known that this non-linear AdS black hole exhibits a vdW's-like phase transition, with the critical point given in Refs. \cite{Du2021,Du2022}
\begin{eqnarray}
r_{c}^{4 \gamma-2}&=&\left(2 q^{2}\right)^{\gamma} f(1, \gamma), \quad f(1, \gamma)=\gamma(4 \gamma-1),~ S_c=\pi \left(2 q^{2}\right)^{\frac{\gamma}{2\gamma-1}} f^{1/(2\gamma-1)}(1, \gamma), \label{Sc}\\
T_{c}&=&\frac{1}{\pi\left(2 q^{2}\right)^{\gamma /(4 \gamma-2)} f^{1 /(4 \gamma-2)}(1, \gamma)} \frac{2 \gamma-1}{4 \gamma-1}, \label{Tc}\\
P_{c}&=&\frac{2 \gamma-1}{16 \pi \gamma\left(2 q^{2}\right)^{\gamma /(2 \gamma-1)} f^{1 /(2 \gamma-1)}(1, \gamma)}.\label{Pc}
\end{eqnarray}
where the YM charge parameter satisfies the condition $\frac{1}{2}<\gamma$. It is clear that this critical point closely depends on the YM charge and the non-linear YM charge parameter. From the above quantities we can obtain a interesting relation, which is only related with the non-linear YM charge parameter, as
\begin{eqnarray}
S_c^2T_c^2P_c=\frac{(2\gamma-1)^3}{16\pi\gamma(4\gamma-1)^2}.
\end{eqnarray}

\begin{figure}[htp]
\centering
\includegraphics[width=0.35\textwidth]{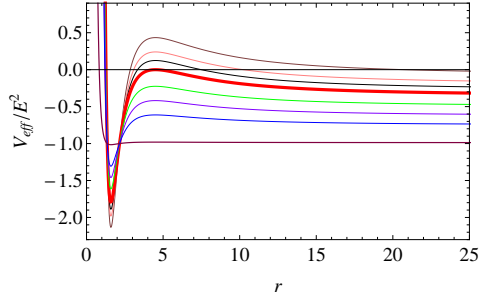}
\caption{The effective potential for the EPYM AdS black h  ole with the parameters $\gamma=1.01,~P=0.003,~q=1.9,~M=2$. The angular momentums $L/E$ of the photon varies from $0.5$ to $36$ from bottom to top. }\label{Veffr}
\end{figure}

Now we will consider a free photon orbiting around a black hole one the equatorial hyperplane defined by $\theta=\pi/2$ and $p_\theta=0$. The Lagrangian takes the form as
\begin{eqnarray}
H=\frac{1}{2}g^{\mu\nu}p_\mu p_\nu=\frac{1}{2}\left(-f^{-1}p_t^2+fp_r^2+r^{-2}p_\phi^2\right),\label{H}
\end{eqnarray}
where $p_\mu$ are the generalized momentums. For this black hole background, there are two Killing fields $\partial_t$ and $\partial_\phi$, which lead two constants, the particle energy $E$ and orbital angular momentum $L$ along each geodesics
\begin{eqnarray}
-E=p_t=-f(r)\dot{t},~~~~~~L=p_\phi=r^2\dot{\phi},
\end{eqnarray}
where the dot represents the derivative to the affine parameter. The light rays are the solutions to Hamilton's equations
\begin{eqnarray}
\dot{p}_\mu=-\frac{\partial H}{\partial x^\mu},~~~~\dot{x}^\mu=\frac{\partial H}{\partial p_\mu},
\end{eqnarray}
which read
\begin{eqnarray}
\dot{p}_t&=&0,~~\dot{p}_\phi=0,~~\dot{p}_\theta=0,\nonumber\\
\dot{p}_r&=&-\frac{1}{2}\left(\frac{f'(r)p _t^2}{f^2(r)}+f'(r)p_r^2+2f(r)p_rp_r'-\frac{2p_\phi^2}{r^3}\right),\nonumber\\
\dot{t}&=&-\frac{p_t}{f(r)},~~\dot{r}=f(r)p_r,~~\dot{\phi}=\frac{p_\phi}{r^2}.\label{dot}
\end{eqnarray}
With $H=0$, we have
\begin{eqnarray}
\dot{r}^2+V_{eff}=0,~~~V_{eff}=\frac{L^2f(r)}{r^2}-E^2.
\end{eqnarray}
As an example, we exhibit the effective potential in Fig. \ref{Veffr} for the EPYM AdS black hole with the fixed parameters $\gamma=1.01,~P=0.003,~q=1.9,~M=2$, and different angular momentums $L/E$. Since the positive of $\dot{r}^2$, there exists the condition: $V_{eff}<0$. Thus the photon can only survive in the range of negative effective potential. When the conserved quantum $L$ is small, the photon will fall into the black hole from somewhere with a larger value of $r$. While, for the larger value of $L$, the peak of effective potential will increase, that leads to the reflection of photon before it falls into black hole. Between these two cases, there exists a critical case which is described by the thickness red line, whose peak approaches zero at $r=4.3122$ and at the same time the radial velocity of photon vanishes. This point just corresponds to the photon sphere radius because of the spherically symmetric static black hole. In the following, we will present the relation between a spatial static observer and the photon sphere radius.

From the Fig. \ref{Veffr}, we knows that the unstable circular photon sphere satisfies
\begin{eqnarray}
V_{eff}=0,~~\frac{dV_{eff}}{dr}=0,~~\frac{d^2V_{eff}}{dr^2}<0.
\end{eqnarray}
Here we denote the photon sphere radius as $r_{ph}$. From the above second equation, we find that the radius $r_{ph}$ satisfies
\begin{eqnarray}
2f(r)\mid_{r_{ph}}=rf'(r)\mid_{r_{ph}},\label{rph}
\end{eqnarray}
where the prime represents the derivative to the radial coordinate $r$. For given the metric function, by solving the equation $V_{eff}=0$, the impact parameter or the angular momentum of the photon reads
\begin{eqnarray}
\mu_{ph}\equiv\frac{L}{E}=\frac{r}{\sqrt{f(r)}}\bigg|_{r_{ph}}.
\end{eqnarray}

Considering light rays sent from an observer at radius coordinate $r_0$ into the past. These light rays can be divided into two classes: Light rays of the first class go to infinity after being deflected by the black hole. Light rays of the second one go towards the horizon of the black hole. If there are no light sources between the observer and the black hole, initial directions of the second class correspond to darkness on the observer's sky. This dark circular disk on the observer's sky is called the shadow of the black hole. The boundary of the shadow is determined by the initial directions of light rays that asymptotically spiral towards the outermost photon sphere. The light ray send from a static observer at position $r_0$ transmits into the past with an angle $\alpha$ relative to the radial direction, which reads
\begin{eqnarray}
\cot{\alpha}=\sqrt{\frac{1}{r^2 f(r)}}\frac{dr}{d\phi}\bigg|_{r_0}.
\end{eqnarray}
If the light ray goes out again after reaching $r_{ph}$, from eq. (\ref{dot}) the orbit equation can be written as
\begin{eqnarray}
\frac{dr}{d\phi}=\pm r\sqrt{\frac{r^2}{\mu_{ph}^2}-f(r)}.
\end{eqnarray}
Thus, the angular radius of the shadow becomes
\begin{eqnarray}
\cot{\alpha}^2=\frac{r_0^2}{\mu_{ph}^2 f(r_0)}-1,
\end{eqnarray}
and by using a trigonometric identity,
$1+\cot{\alpha}^2=\frac{1}{\sin^2{\alpha}}$, it can be rewritten as
\begin{eqnarray}
\sin{\alpha}=\frac{\mu_{ph}\sqrt{f(r_0)}}{r_0}
\end{eqnarray}
The shadow radius of the black hole observed by a static observer at $r_0$ can be written as
\begin{eqnarray}
r_s
=\frac{r_{ph}}{\sqrt{f(r_{ph})}}\sqrt{f(r_0)}.\label{rs}
\end{eqnarray}

\section{Phase transition of EPYM AdS black hole from the viewpoint of shadow radius}
\label{scheme3}
When this system is undergoing the first-order phase transition (HPBH/LPBH phase transition) with temperature $T$ ($T=\chi T_c,~\chi<1$), we denote the horizon radius of two coexistence black hole phases as $r_1$ and $r_2$, respectively. With the Maxwell's equal area law, the condition of the first-order phase transition for this system with the given temperature $T$, which had been shown in our previous works \cite{Du2021,Du2022}, reads
\begin{eqnarray}
\chi\frac{2\gamma-1}{\gamma^{1/(4\gamma-2)}(4\gamma-1)^{(4\gamma-1)/(4\gamma-2)}}
=\frac{1}{f^{1/(4\gamma-2)}(x,\gamma)}\left(1+x-
\frac{1-x^{4\gamma}}{2f(x,\gamma)(1-x)x^{4\gamma-2}}\right),~~
\frac{(2q^2)^\gamma}{r_2^{4\gamma-2}}=\frac{1}{f(x,\gamma)} \label{condition}
\end{eqnarray}
with
\begin{eqnarray}
f(x,\gamma)=\frac{(3-4\gamma)(1+x)(1-x^{4\gamma})+8\gamma x^2(1-x^{4\gamma-3})}{2x^{4\gamma-2}(3-4\gamma)(1-x)^3},~~x=\frac{r_{1}}{r_{2}}.
\end{eqnarray}

\begin{figure}[htp]
\centering
\subfigure[$\gamma=1$]{\includegraphics[width=0.3\textwidth]{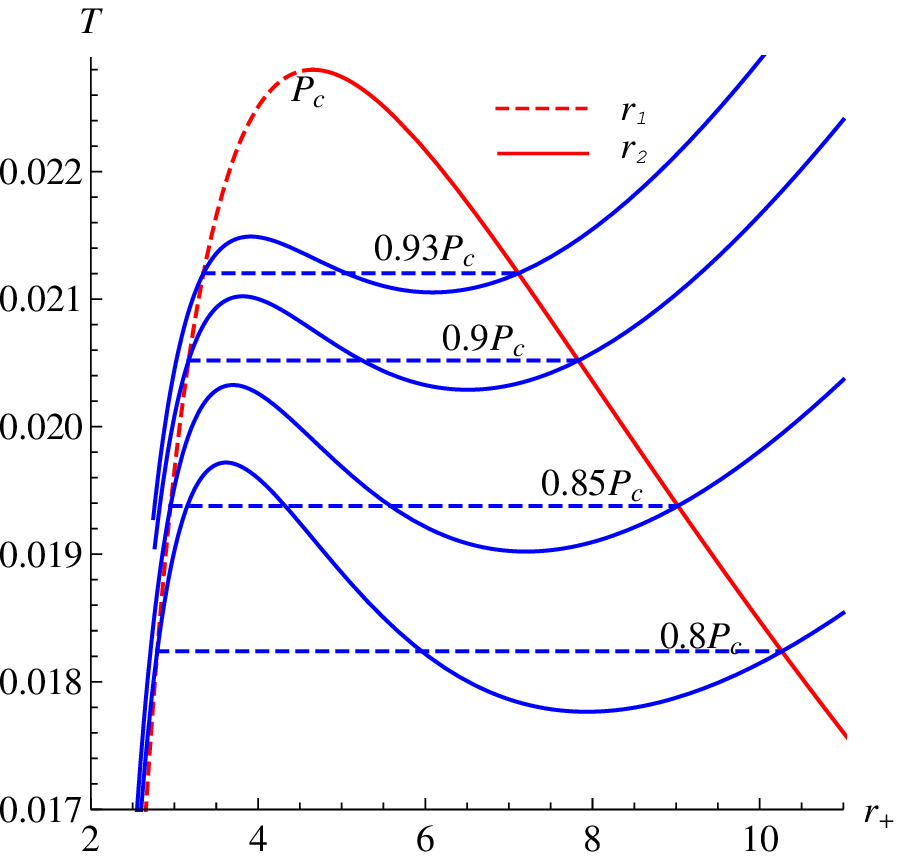}}~~
\subfigure[$\gamma=1$]{\includegraphics[width=0.3\textwidth]{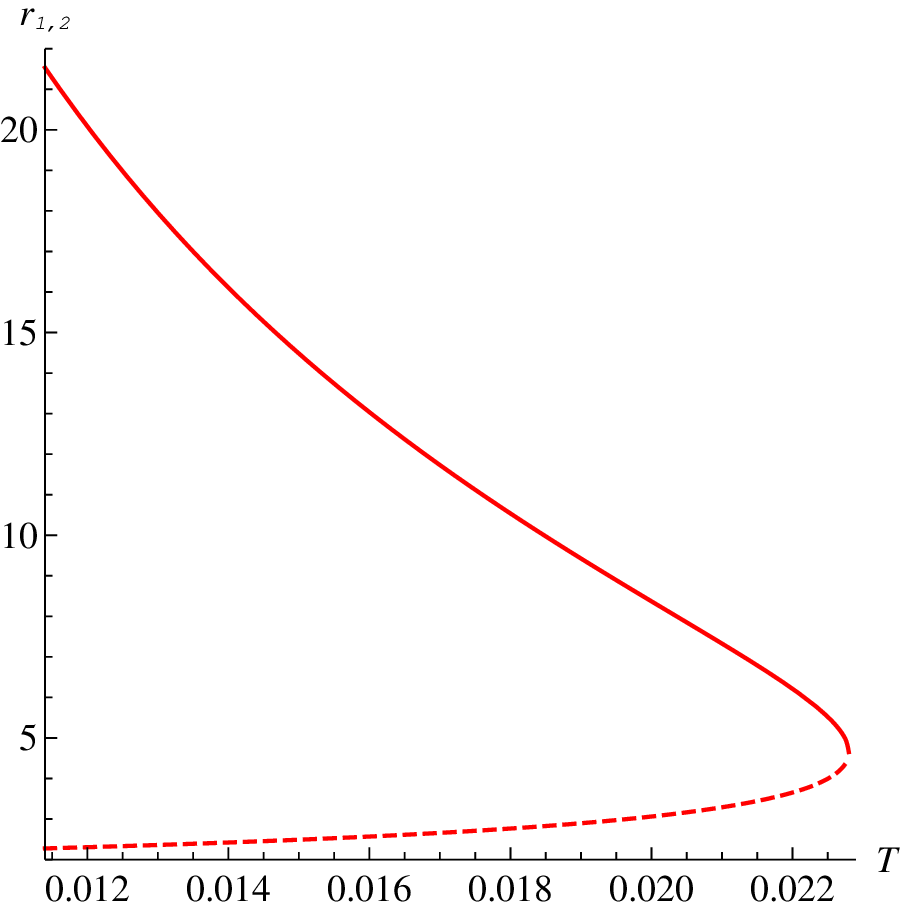}}~~
\subfigure[$\gamma=1.5$]{\includegraphics[width=0.3\textwidth]{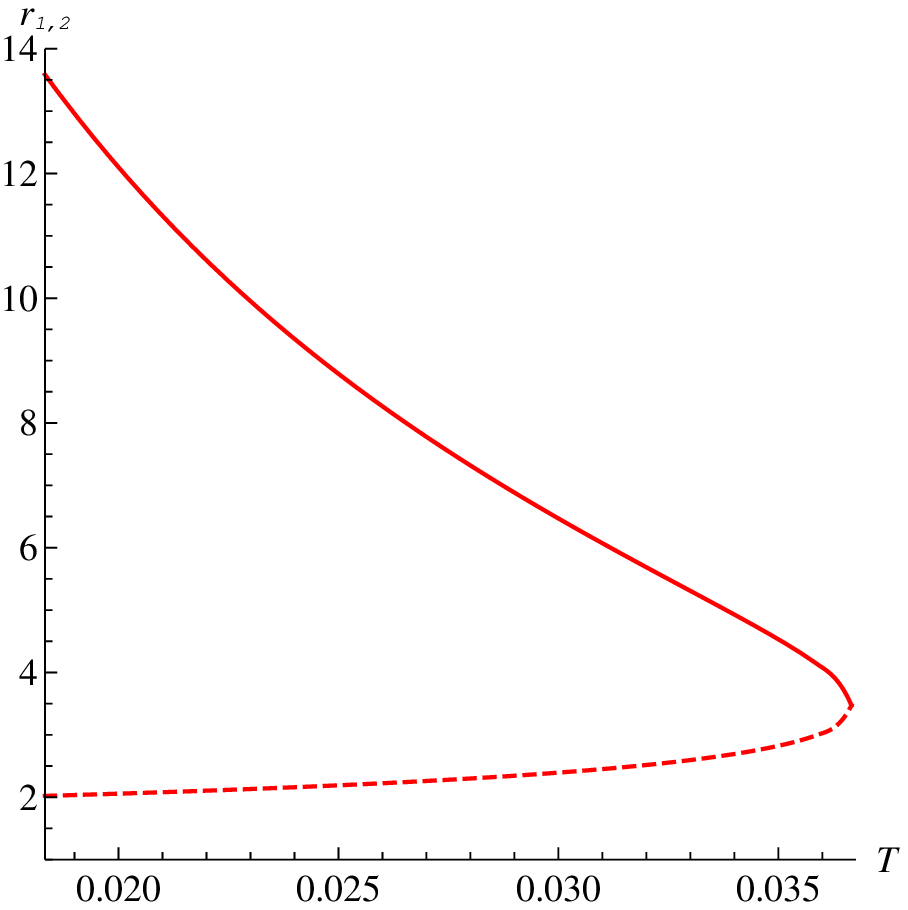}}
\caption{The pictures of the horizon radii for two coexistent black hole phases with temperature. The YM charge set to $q=1.9$. }\label{r11912-T}
\end{figure}

As $q=1.9$, by numerically fitting the data the horizon radii of the coexistent big and small black hole phases nearby the critical point as the functions of temperature are given as the following
\begin{eqnarray}
r_1&\approx&
\begin{cases}
2.65\times10^7T^4-8.16\times10^5T^3+8.77\times10^3T^2+1.9,  & ~~~~~~~~~~\gamma=1\\
2.39\times10^6T^4-1.2\times10^5T^3+2.16\times10^3T^2+1.77, & ~~~~~~~~~~\gamma=1.5
\end{cases}\\
r_2&\approx&
\begin{cases}
-7.11+3.81\times10^{-10}/T^3-2.52\times10^{-6}/T^2+0.32/T,  &  ~~~~~~\gamma=1 \\
-4.12+8.1\times10^{-10}/T^3-3.27\times10^{-6}/T^2+0.32/T. &  ~~~~~~\gamma=1.5
\end{cases}
\end{eqnarray}
And the coexistent curves read
\begin{eqnarray}
P\approx
\begin{cases}
-6.42\times10^5T^8+6.33\times10^3T^7-1.49\times10^3T^6+7.27T^5+1.13\times10^3T^4-5.58T^3+1.3T^2,  & \gamma=1 \\
250T^8+17.5T^7-116.9T^6-16.4T^5-0.57T^4+15.32T^3+1.01T^2. & \gamma=1.5
\end{cases}
\end{eqnarray}
Nearby the critical point, the radii of two coexistent black hole phases as the function of temperature are exhibited in Fig. \ref{r11912-T}. It is obviously that the horizon radius for the small coexistent phase increases with the increasing of temperature, while for the large coexistent phase it increases with temperature.

Now, we will investigate the phase transition information of this
system from the viewpoint of the shadow radius. In order to obtain the analyzed result, firstly we set the non-liner YM charge parameter equal to one: $\gamma=1$. The photon sphere radius becomes
\begin{eqnarray}
r_{ph}=\frac{1}{2}\left(3M+\sqrt{9M^2-8q^2}\right).
\end{eqnarray}
It is clear that this result is exactly equal to that of the asymptotically flat charged black hole, in which the photon sphere radius depends on the pressure. Note that the mass of eq. (\ref{M}) is related with the pressure. Substituting it into above equation, we have the photon sphere radius
\begin{eqnarray}
r_{ph}
=\frac{3q^2+r_+^2(3+8\pi Pr_+^2)+\sqrt{[3q^2+r_+^2(3+8\pi Pr_+^2)]^2-32 q^2r_+^2}}{4r_+}.\label{rph1}
\end{eqnarray}
At the critical point ($r_+=r_c$ and $P=P_c$), the critical photon sphere radius reads $r_{phc}=(2+\sqrt{6})q$ for $\gamma=1$. Therefore the shadow radius as a function of the black hole horizon radius becomes
\begin{eqnarray}
r_s=\frac{r_{ph}(r_+)}{\sqrt{f(r_{ph}(r_+))}}\sqrt{f(r_0)}.\label{rsnew}
\end{eqnarray}
Substituting the critical photon sphere radius and pressure into eq. (\ref{rsnew}), we can get the critical shadow radius $r_{sc}$, which is related to the YM charge $q$.

\begin{figure}[htp]
\centering
\subfigure[$~q=1.9$]{\includegraphics[width=0.3\textwidth]{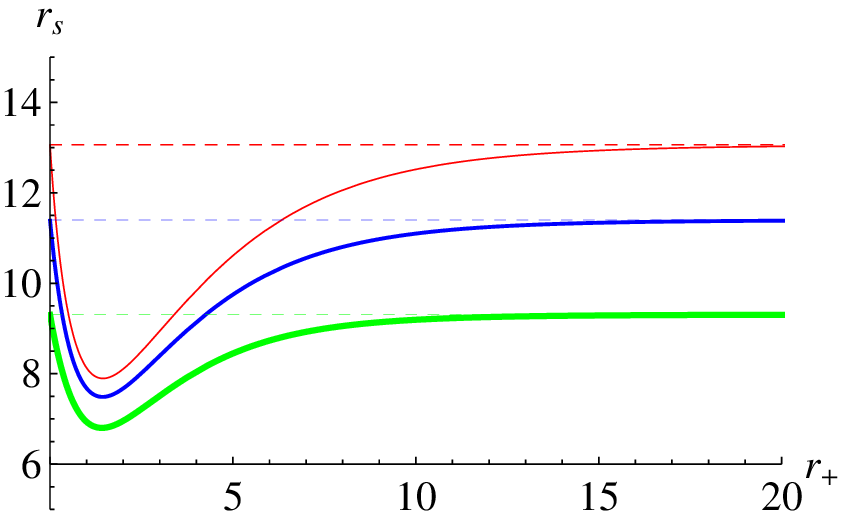}\label{rh-rs0}}~
\subfigure[$~P=0.0006994<P_c$]{\includegraphics[width=0.3\textwidth]{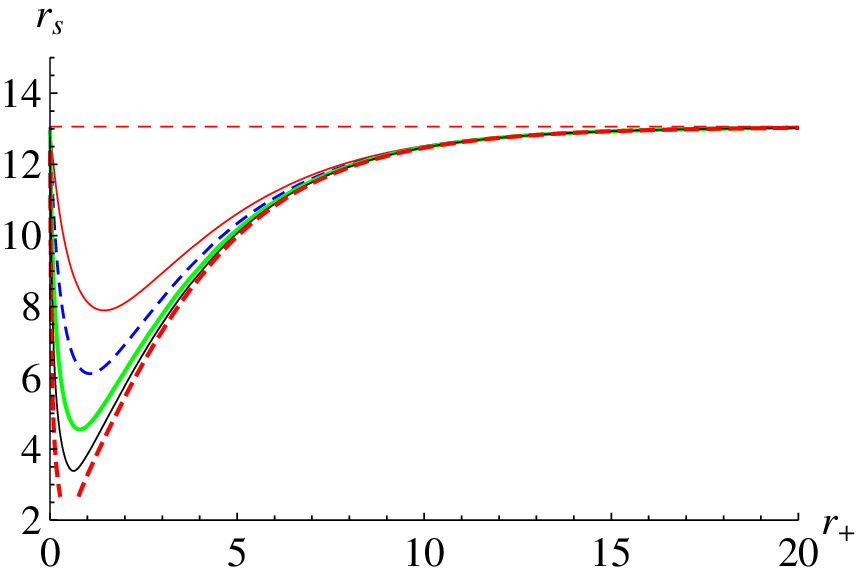}\label{rh-rs-q}}~
\subfigure[$~P=1.5P_c$]{\includegraphics[width=0.3\textwidth]{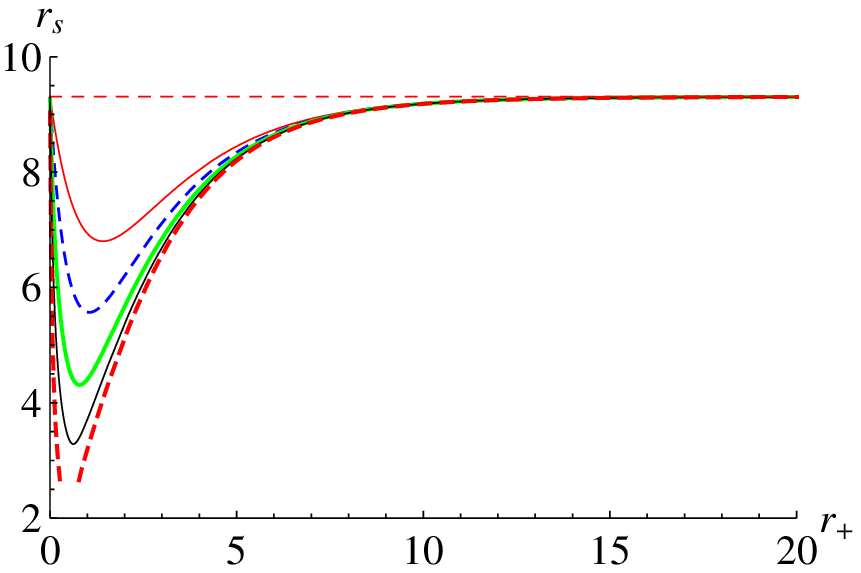}\label{rh-rs-q0}}~
\caption{The non-linear YM charge parameter set to $\gamma=1$. In the left picture: the pressure set to $P_c-0.00022,~Pc,$ and $1.5P_c$ from bottom to top. In the middle and right pictures: the YM charge varies from $1.9$ to $0.8$ from top to bottom. The horizontal thin dashed lines stand for the limited value of the shadow radius. }
\end{figure}

Based on the constraint, a static observer at spatial infinity have $f(r_0)=1$ \cite{Zhang2020}. With eqs. (\ref{rph1}) and (\ref{rsnew}), the behaviors of the shadow radius as the function of the black hole horizon radius for different values of pressure can be obtained, which are exhibited in Fig. \ref{rh-rs0}. It is obviously that there exists a minimum value of the shadow radius for this system with the given pressure (no matter it is large than or less than the critical pressure). The shadow radius is decreasing with the increasing of the horizon radius for small black holes, while for the big ones it is increasing with the horizon radius until to a constant $\sqrt{3/(8\pi P)}$. It indicates that at the minimum point the system will undergo a phase transition which is different the HPBH/LPBH one. Furthermore, for the small EPYM black hole its shadow radius decreases with the increasing of pressure.

\begin{figure}[htp]
\centering
\subfigure[$~P=P_c-0.00022,~P_c,~P_c+0.0002$]{\includegraphics[width=0.3\textwidth]{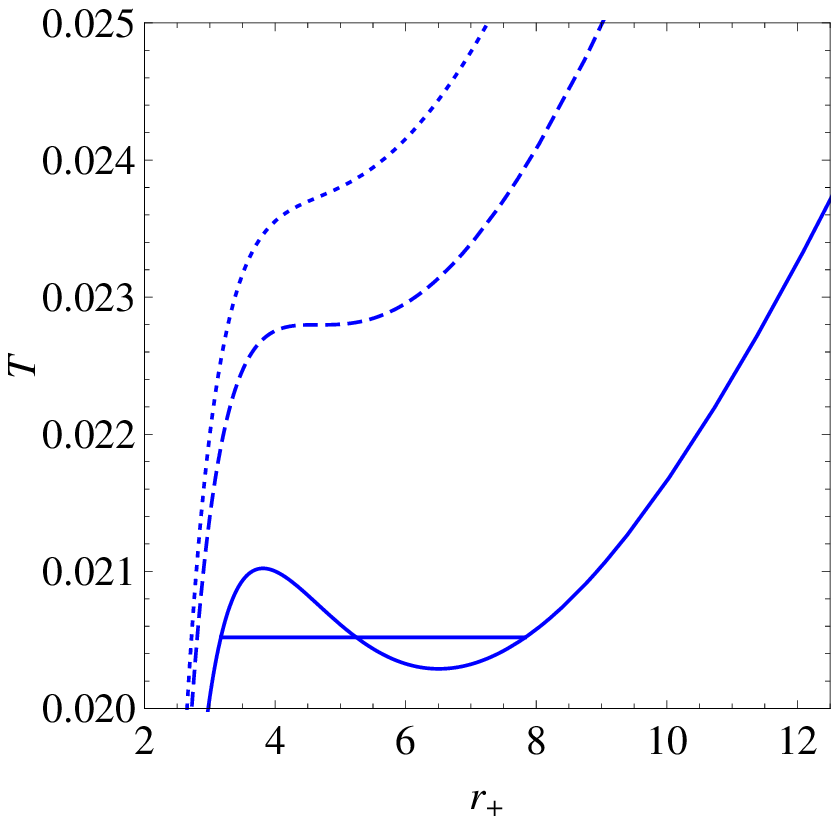}\label{T-rP}}~~~
\subfigure[$~P=P_c-0.00022,~P_c,~P_c+0.0002$]{\includegraphics[width=0.3\textwidth]{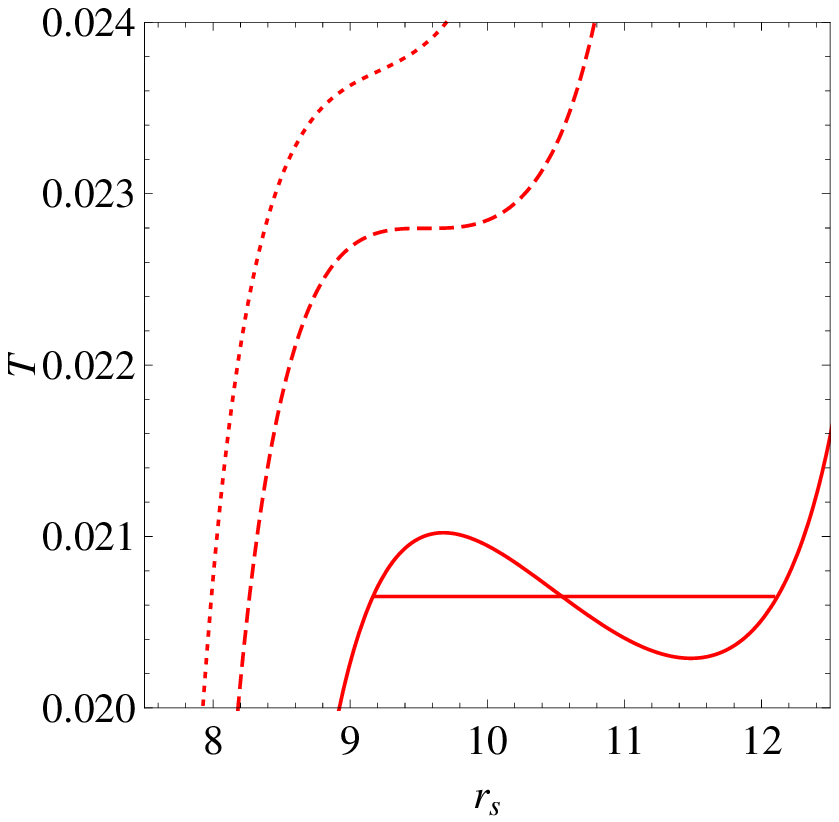}\label{T-rsP}}~~~
\subfigure[$~P=P_c-0.00022$]{\includegraphics[width=0.3\textwidth]{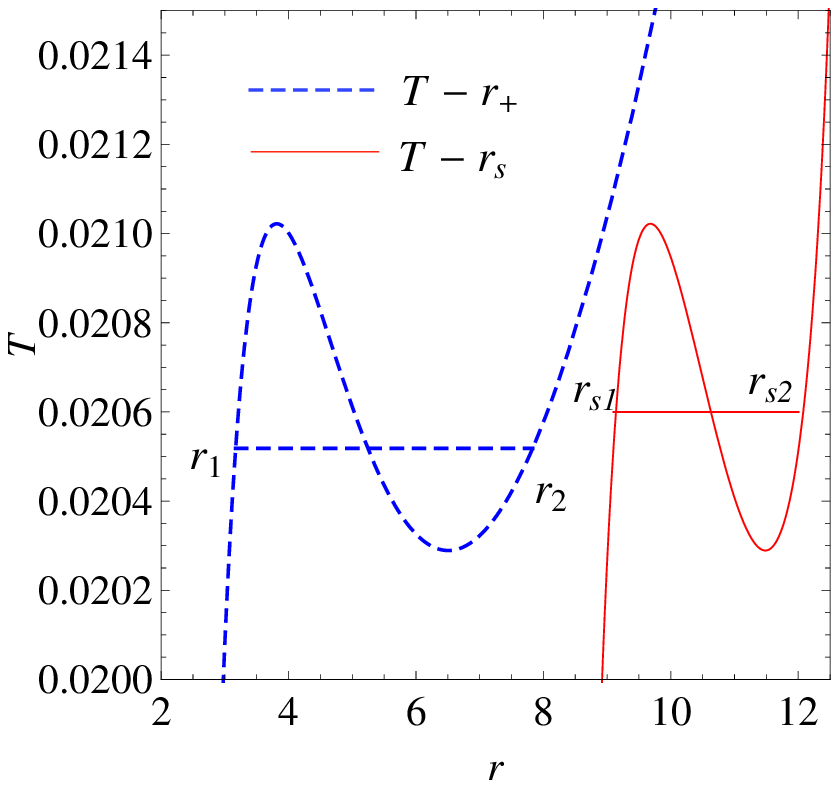}\label{T-rs-rP}}
\caption{The parameters set to $\gamma=1,~q=1.9$. A static observer at $r_0=100$.}\label{T-rs-r}
\end{figure}

There is a interesting phenomena: the values of the shadow radius under the limitations of $r_+\rightarrow0$ and $r_+\rightarrow\infty$ are the same, which is related to the pressure $P$, but not to the YM charge $q$ (shown in Figs. \ref{rh-rs-q} and \ref{rh-rs-q0}). Here we present the limited forms of the shadow radius with different $\gamma$:
\begin{eqnarray}
r_s|_{r_+\rightarrow0,\infty}=
\begin{cases}
\sqrt{-\frac{3}{\Lambda}}=\sqrt{\frac{3}{8\pi P}},  & \gamma=1\label{rsl1} \\
\frac{4.3416}{4\pi\sqrt{P}}, & \gamma=1.5\label{rsl2}
\end{cases}
\end{eqnarray}
Note that for two cases of different non-linear YM charge parameter values, the inverse square coefficient between $r_s|_{r_+\rightarrow0,\infty}$ and pressure is the same: $0.345494$. That means for the EPYM AdS black hole the limited value of the shadow radius is independent with the charge information ($q$ and $\gamma$) carried by the system, only is related with the pressure.

For different values of pressure, combining the first-order phase transition condition (\ref{condition}) and eqs. (\ref{fr}), (\ref{T}), (\ref{Pc}), (\ref{rph1}), and (\ref{rsnew}), we show the $T-r_{+,s}$ phase diagrams in Figs. \ref{T-rP} and \ref{T-rsP}. It is obviously that for $P<P_c$, there exists the non-monotonic behavior both in the $T-r_+$ and $T-r_s$ planes. However, the temperature becomes only a monotone increasing function of the shadow radius and the horizon radius when the pressure is bigger than the critical one. At the critical pressure, there exist the deflection points at $r_+=r_{c}$ and $r_s=r_{sc}$, respectively. These behaviors are very similar to the isobar process of a VdW's system in the $T-S$ plane, which means there exist the first-order phase transition both from the viewpoint of the horizon radius and the shadow radius. Thought constructing the Maxwell's equal-area law both in $T-r_+$ and $T-r_s$ planes, we also get the phase transition points. The corresponding shadow radii are remarked by $r_{s_1}$ and $r_{s_2}$.

It should be noted that there is a rather peculiar phenomenon: for the given pressure $P<P_c$ the phase transition temperature from the viewpoint of the horizon radius is slightly lower than that from the viewpoint of the shadow radius as $\gamma=1$ shown in Fig. \ref{T-rs-rP}. The reason of this situation may be caused by the gravitational effect and the non-linear YM charge term. So far, we have converted the phase transition information of the EPYM AdS black hole phase transition into the measurable physical quantity, the shadow radius. That means we can detect whether there is a phase transition by measuring the shadow radius of the black hole, so as to obtain the microscopic structure of the black hole.

\begin{figure}[htp]
\centering
\subfigure[$~q=1.9$]{\includegraphics[width=0.3\textwidth]{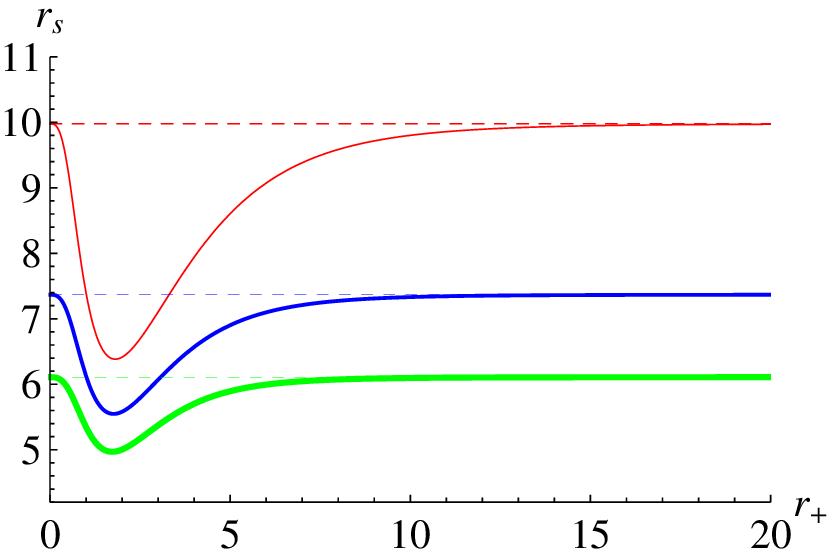}\label{rh-rs15}}~
\subfigure[$~P=0.0012<P_c$]{\includegraphics[width=0.3\textwidth]{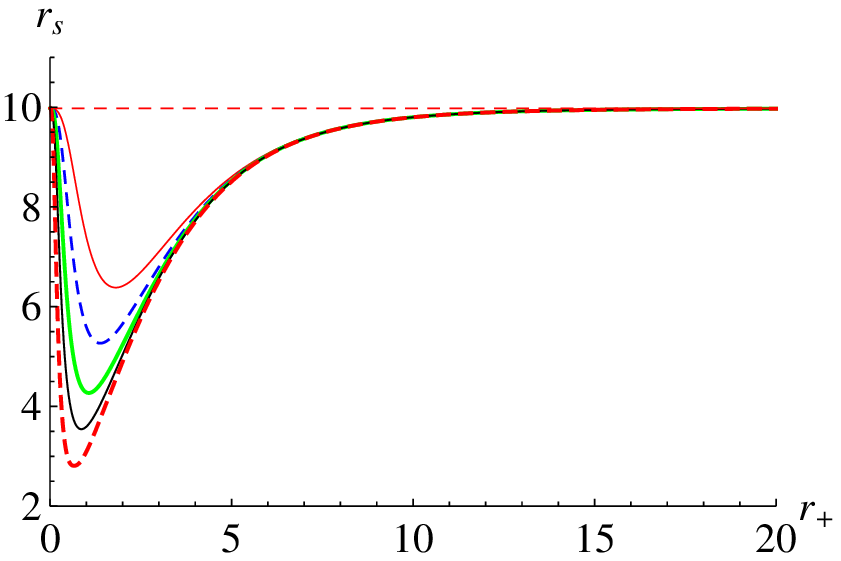}\label{rh-rs-q15}}~
\subfigure[$~P=3.5P_c$]{\includegraphics[width=0.3\textwidth]{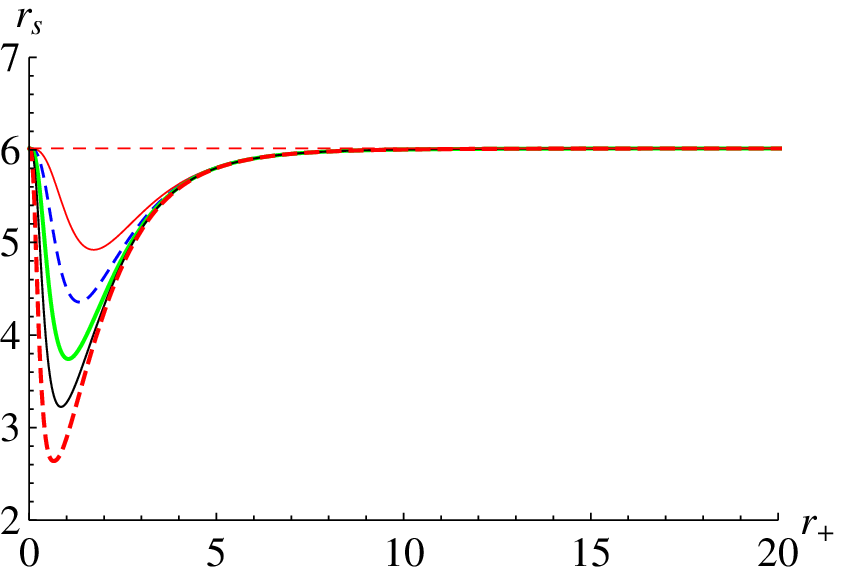}\label{rh-rs-q015}}~
\caption{The non-linear YM charge parameter set to $\gamma=1.5$. In the left picture: the pressure set to $P_c-0.0001,~Pc,$ and $1.5P_c$ from bottom to top. In the middle and right pictures: the YM charge varies from $1.9$ to $1$ from bottom to top. The horizontal thin dashed lines stand for the limited value of the shadow radius.}
\end{figure}

\begin{figure}[htp]
\centering
\subfigure[$~P=P_c-0.0005,~P_c,~P_c+0.0005$]{\includegraphics[width=0.3\textwidth]{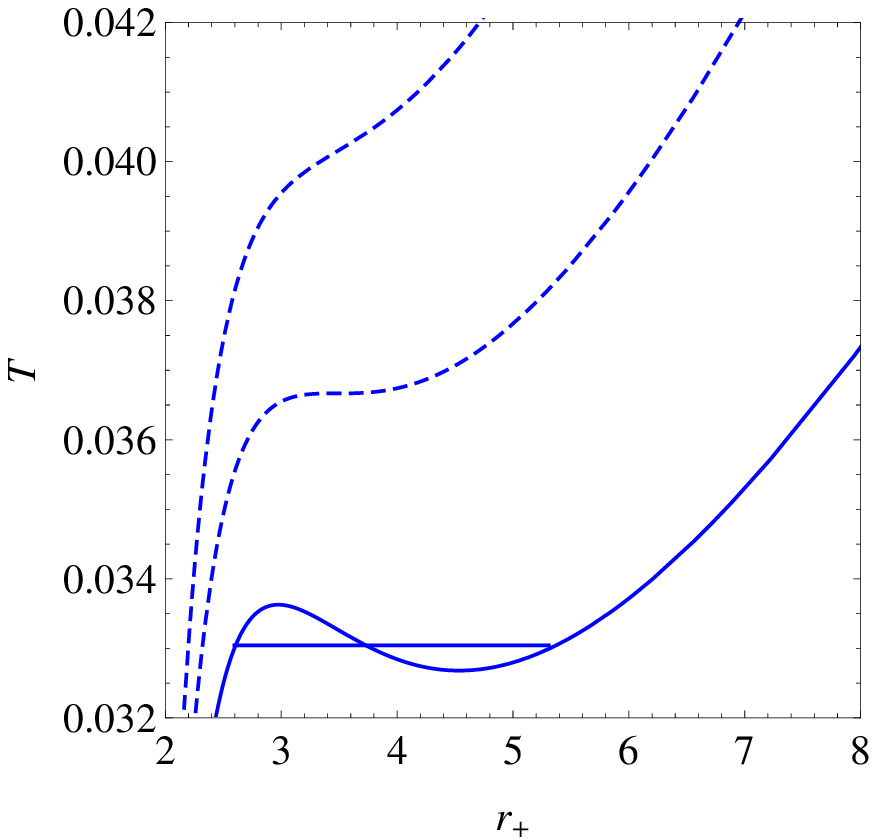}\label{T-rP15}}~~~
\subfigure[$~P=P_c-0.0005,~P_c,~P_c+0.0005$]{\includegraphics[width=0.3\textwidth]{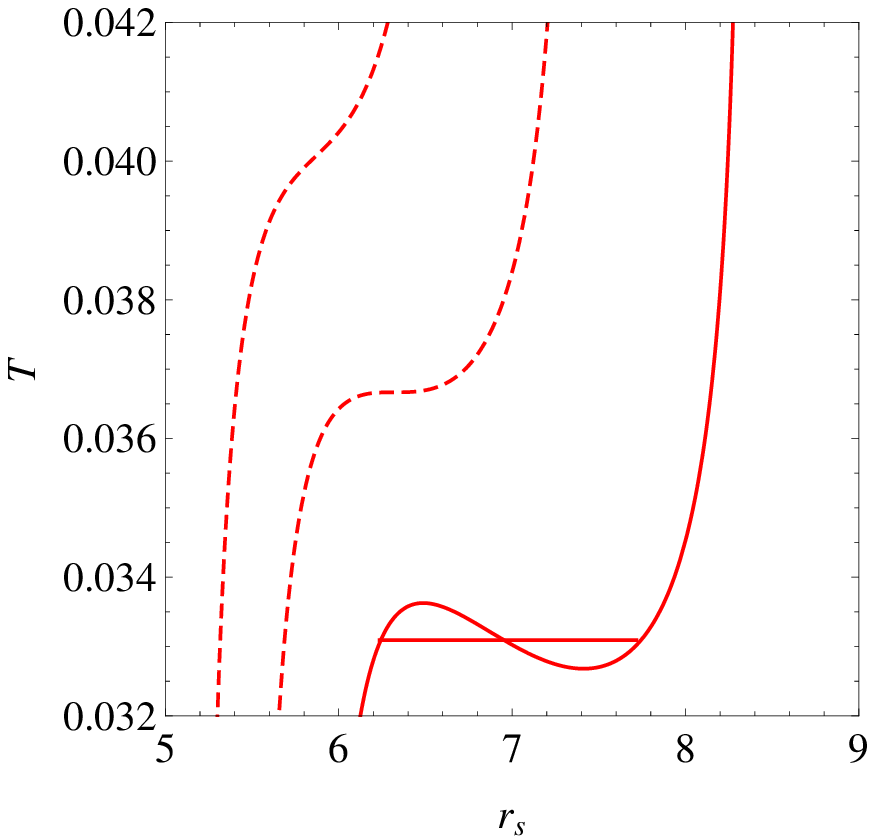}\label{T-rsP15}}~~~
\subfigure[$~P=P_c-0.0005$]{\includegraphics[width=0.3\textwidth]{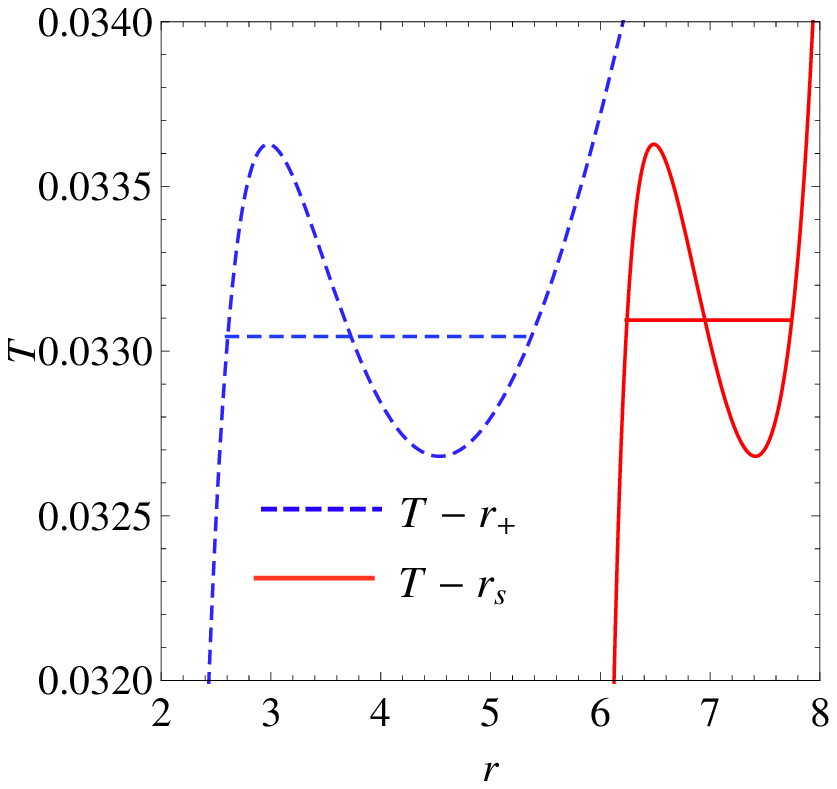}\label{T-rs-rP15}}
\caption{The parameters set to $\gamma=1.5,~q=1.9$. A static observer at $r_0=100$.}\label{T-rs-r15}
\end{figure}

In addition, the non-monotonic behavior of $T-r_s$ only just indicates the existence of the first-order phase transition, but not the first-order one between high potential black hole and low potential black hole. Since the black hole horizon radius cannot be directly obtained from experiments, while its shadow can be observed. Therefore, the shadow radius may serve as a probe for the phase structure of the EPYM AdS black hole. For the case of $\gamma=1.5$ we also present the numerical results of the shadow radius with the black hole horizon radius for different pressures and YM charges in Fig. \ref{rh-rs15}, whose behaviors are similar to the case of $\gamma=1$. In this case, the limited form of the shadow radius is exhibited in eq. (\ref{rsl2}). The behaviours for the temperature as the functions of the horizon radius and the shadow radius have been shown in Fig. \ref{T-rs-r15}, which are similar to that in the case of $\gamma=1$.

\begin{figure}[htp]
\centering
\subfigure[$\gamma=1$]{\includegraphics[width=0.3\textwidth]{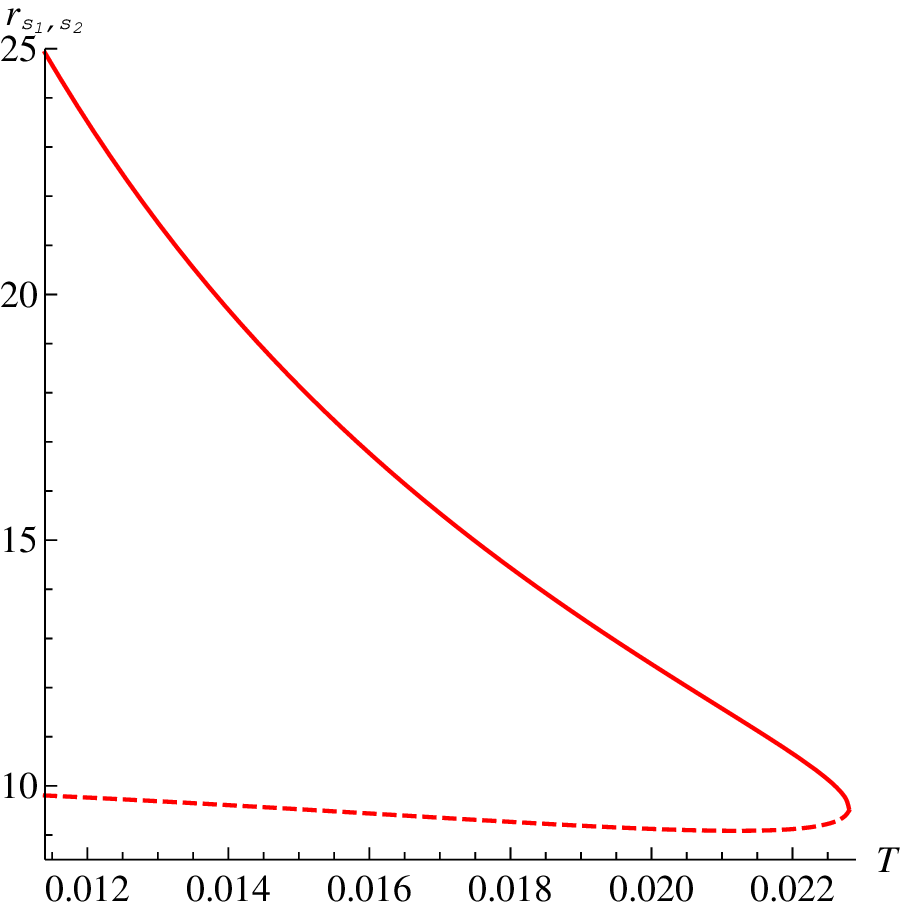}\label{rs12-T}}~~~~~~~
\subfigure[$\gamma=1$]{\includegraphics[width=0.3\textwidth]{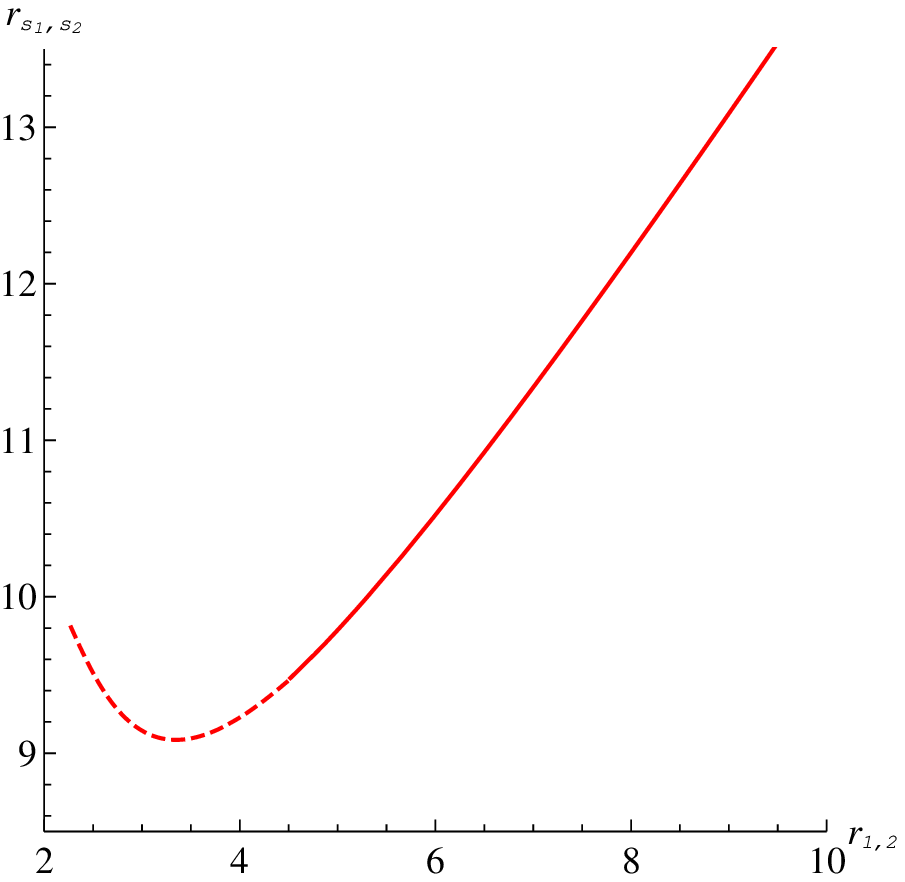}\label{rs12-r12}}\\
\subfigure[$\gamma=1.5$]{\includegraphics[width=0.3\textwidth]{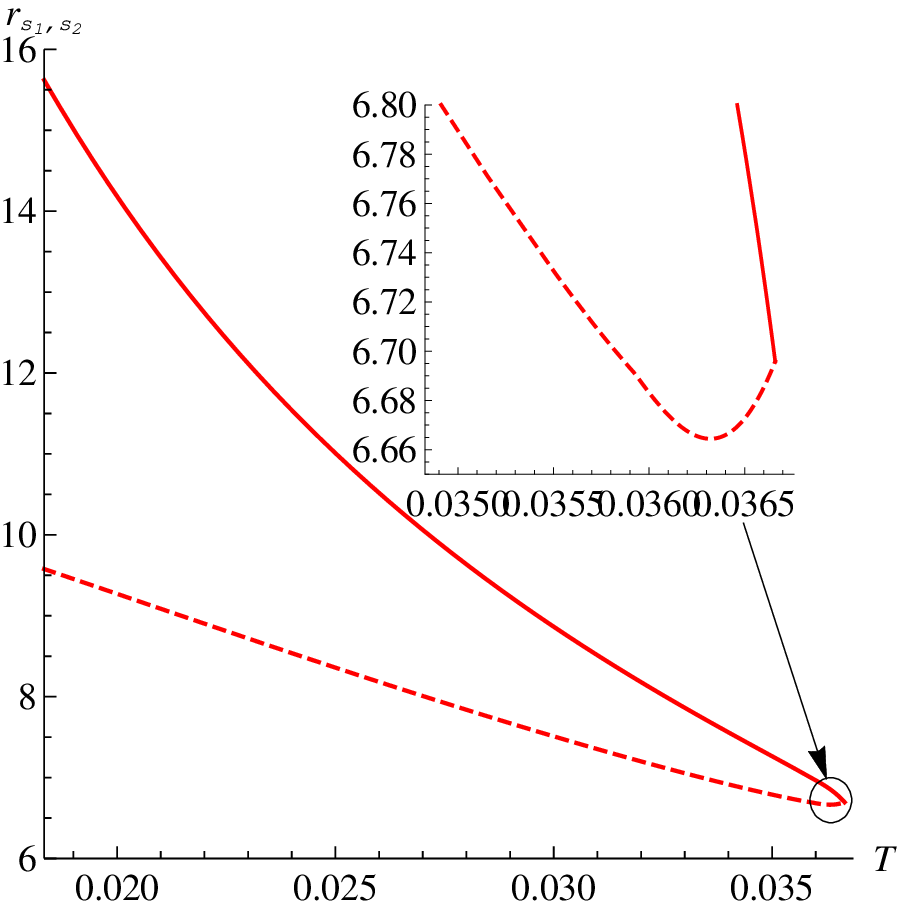}\label{rs12-T15}}~~~~~~~
\subfigure[$\gamma=1.5$]{\includegraphics[width=0.3\textwidth]{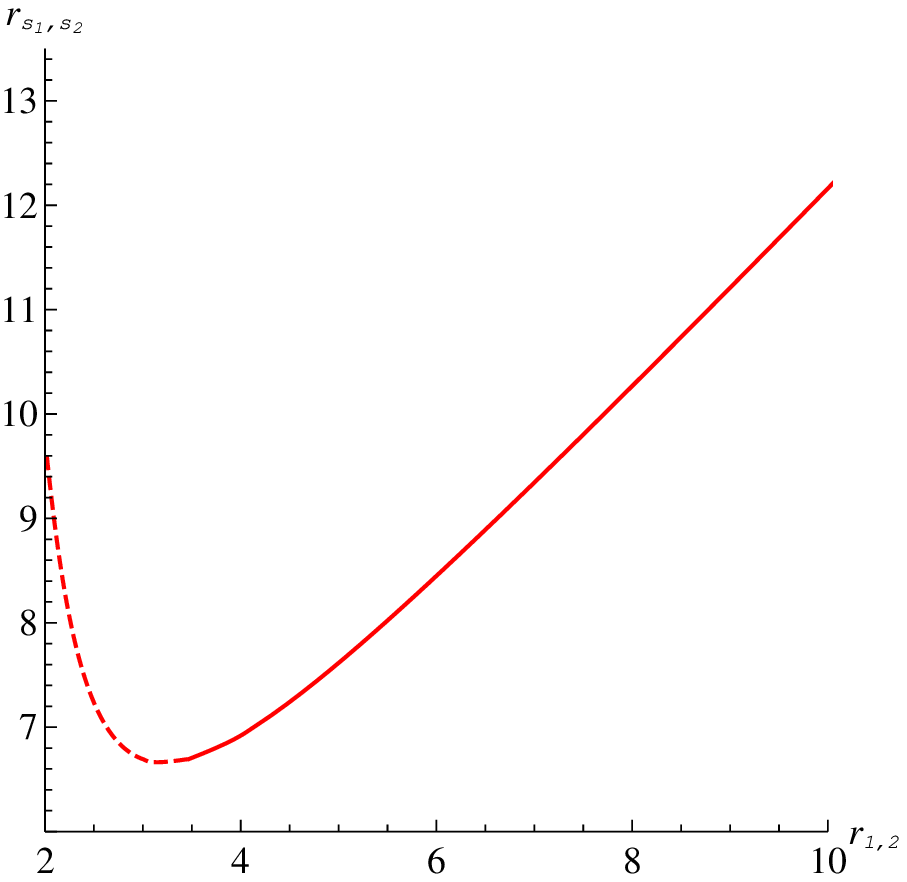}\label{rs12-r12-15}}
\caption{The parameters set to $q=1.9$. And the temperature various from $0.5T_c$ to $Tc$. A static observer at $r_0=100$.}\label{rs12-T-r12}
\end{figure}

In order to exhibit the relationship between the black hole phase transition and the shadow radius in two cases of $\gamma=1,1.5$, the pictures of the shadow radii for two-coexistent black hole phases as the functions of temperature and horizon radius are shown in Fig. \ref{rs12-T-r12}. For the big coexistent phase, its shadow radius decreases with the increasing of temperature, while it increases with the horizon radius. However for the small one its shadow radii as the function of temperature and the horizon radius, both have a local minimum value. At the local minimum point there maybe exists a new phase transition.

\section{Thermal profile of the EPYM AdS BH}
\label{scheme4}

\begin{figure}[htp]
\centering
\subfigure[$~r_+=r_2$]{\includegraphics[width=0.3\textwidth]{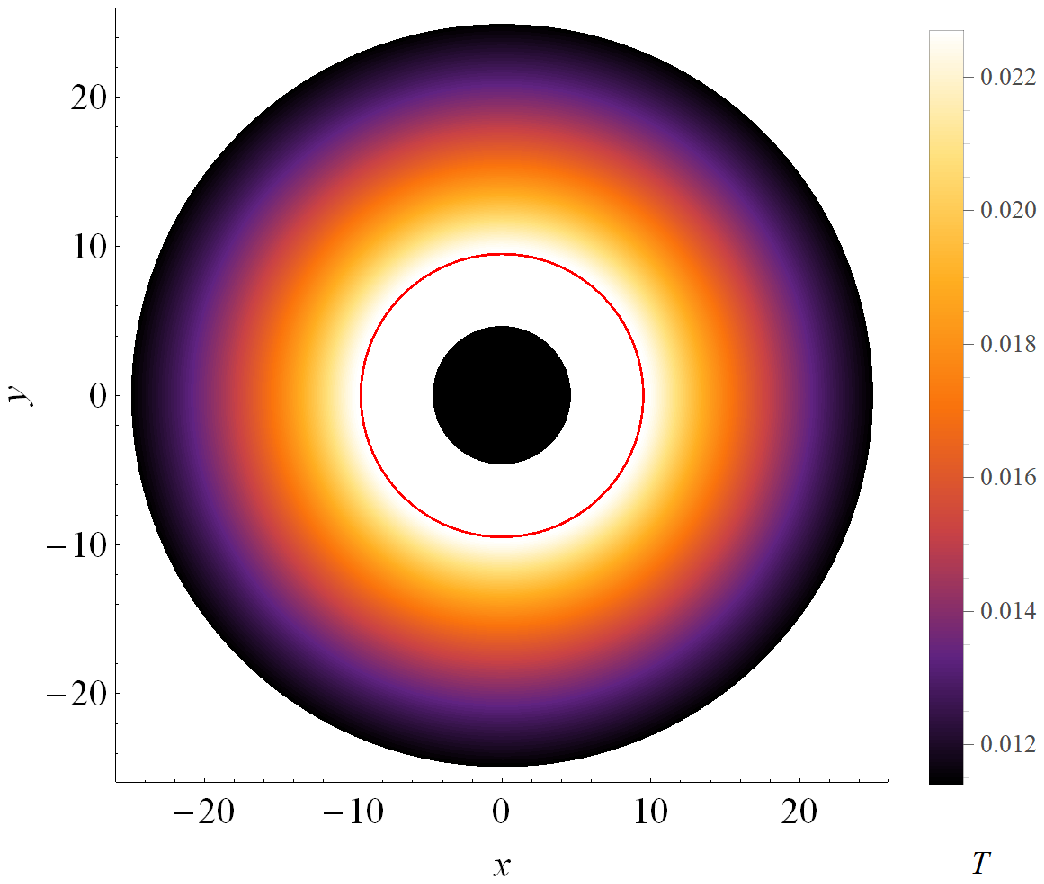}\label{xy10}}~~~~~~
\subfigure[$~r_+=r_1$]{\includegraphics[width=0.3\textwidth]{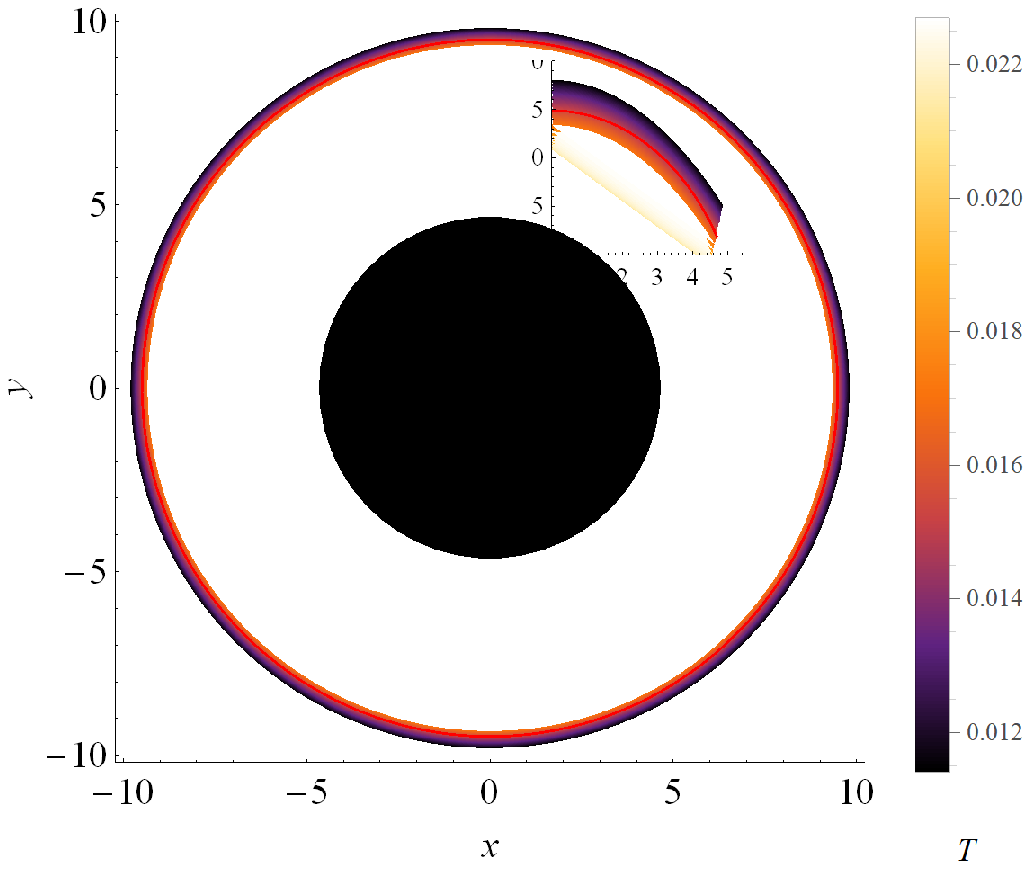}\label{xy20}}\\
\subfigure[$~r_+=r_2$]{\includegraphics[width=0.3\textwidth]{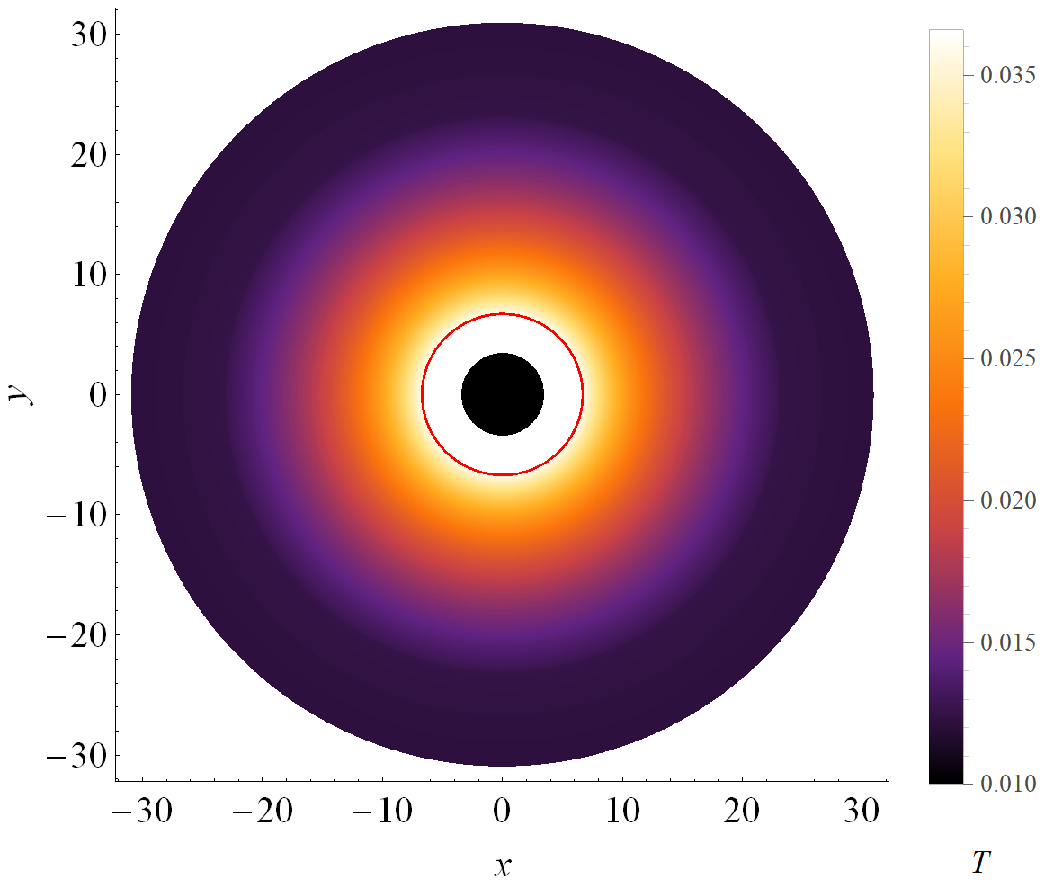}\label{xy115}}~~~~~~
\subfigure[$~r_+=r_1$]{\includegraphics[width=0.3\textwidth]{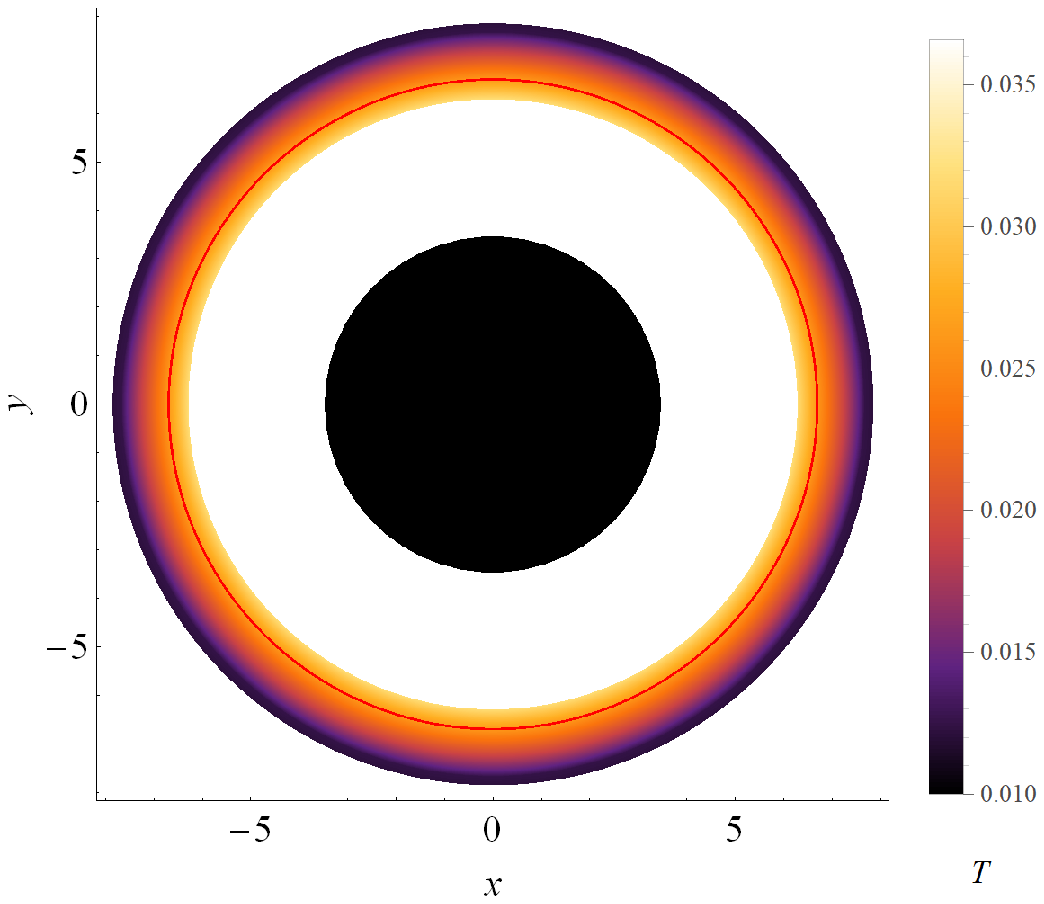}\label{xy215}}
\caption{The shadow pictures of the coexistent big and small black hole phases with temperature. The parameters set to $q=1.9$ and $\gamma=1$ (in Figs. \ref{xy10} and \ref{xy20}), $\gamma=1.5$ (in Figs. \ref{xy115} and \ref{xy215}). The temperature various from $0.5T_c$ to $T_c$. }\label{xy}
\end{figure}

Since the shape of a spherically symmetric black hole shadow is circular for any observer \cite{Chang2020}, we will establish a thermal profile in the two-dimensional plane to more intuitively present the relationship between the BH phase structure and its shadow for the EPYM AdS black hole. According to Ref. \cite{Eiroa2019}, the shadow boundary curve at the celestial coordinate reads as
\begin{eqnarray}
x=\lim_{r\rightarrow\infty}\left(-r^2\sin\theta_0\frac{d\phi}{dr}\right)_{\theta_0\rightarrow\pi/2},~~
y=\lim_{r\rightarrow\infty}\left(r^2\frac{d\theta}{dr}\right)_{\theta_0\rightarrow\pi/2}.
\end{eqnarray}
As $\gamma=1$ and $1.5$, the shadow contours of the coexistent big and small black hole phases with temperature for a static spatial infinity observer are exhibited in Fig. \ref{xy}.

The results show that the size of the BH shadow depends on the temperature. When $T<T_c$, the shadow radius for the coexistent big black hole phase decreases monotonically with temperature until to the critical shadow and it is in the large radius region, which correspond to the supercritical black hole phases; the red thick curves stand for the critical shadow; for the small black hole phases the shadow radii support in the small radius region; the inner black disks stand for the coexistent big and small black hole phases with different temperatures. It is obviously that for the coexistent small phases, the shadow radius decreases gently with the lower temperature until to the minimum value, then increases drastically with the higher temperature. These characters are consistent with those exhibited in Fig. \ref{rs12-T-r12}.


\section{Discussions and Conclusions}
\label{scheme5}
In this manuscript, we examined the relationship between the thermodynamic phase transition for the four-dimensional charge Einstein-power-Yang-Mills (EPYM) AdS black hole and its shadow. This will provide a new way to realize the link between the gravity and thermodynamics of black holes.

First, we presented the characters of EPYM AdS black hole phase transition: the critical point depend on the YM charge and non-linear YM charge parameter with the condition $\gamma>1/2$ and $\gamma\neq3/4$. Since the critical point stands for the boundary of the coexistent phases, it is keys to probe the black hole phase structure. Then we investigated the null geodesics of a photon in the equatorial plane of EPYM AdS black hole background. By analyzing the effective potential of photon orbits with the certain parameters, we obtained the photon sphere radius and impact parameter (angular momentum of the photon sphere). For an  infinity static observer, the corresponding black hole shadow can be expressed as the function of the photon sphere radius directly, namely the shadow is related with the black hole horizon. The results showed that the non-monotonic behaviors appear in $T-r_s$ and $r_s-r_{+}$ diagrams. That indicated that there exists a certain relationship between the shadow and the phase transition of this system. In addition, we also presented the influence of non-linear YM charge parameter $\gamma$ on the shadow.

Finally, we further explored the relationship between the shadow, temperature, and two-existent black hole phases. The relevant conclusions can be summarized as follows:
\begin{itemize}
  \item{For a isobar process of this system, the shadow radius as the function of the black hole horizon has the non-monotonic behavior with the certain parameters. In the $r_s-r_+$ plane, the local stable minimum maybe indicate a new phase transition, which is different from with the black hole phase transition. Furthermore, for a given pressure, the limitations of the shadow radius with $r_+\rightarrow0,\infty$ are the same one. That means as $r_+\rightarrow0,\infty$ the limitation of the shadow radius is independent with the charge information ($q$ and $\gamma$) carried by the system, only is related with the pressure;}
  \item{For a given pressure, which is less than the critical value, there exists two extremal points in the $T-r_s$ plane. Until the pressure approaches to the critical one, two extremal points coincide with each other. When the pressure is larger than the critical one, there is no extremal point. Those behaviors of the shadow radius are consistent with that of the black hole phase transition. Therefore, the behavior of photon sphere can be regarded a probe to reveal the thermodynamic phase transition information of black holes;}
  \item{For the same set of parameters, the phase transition temperature from the viewpoint of the shadow radius is slightly higher than that from the viewpoint of the black hole horizon, which is maybe caused by the gravitational effect and the non-linear YM charge term;}
  \item{At the coexistent curve of $T-P$,  the shadow radius for the coexistent big black hole phase decreases monotonically with temperature until to the critical shadow, which correspond to the supercritical black hole phases; for the coexistent small black hole phases, the shadow radius decreases gently with the lower temperature until to the minimum value, then increases drastically with the higher temperature. These characters are consistent no matter in the $r_{s_1,s_2}-r_{1,2}$ planes and the thermal profiles.}
\end{itemize}

Those analyze firmly enhance our conjecture that there exists the relationship between the null geometry and thermodynamic phase transition for EPYM AdS black holes. It further supports the link between the gravity and thermodynamics of black holes, and also provides a possible way to describe the strong gravitational effect from the thermodynamic side.

\section*{Acknowledgements}

We would like to thank Prof. Ren Zhao, Meng-Sen Ma, and Yu-Peng Zhang for their indispensable discussions and comments. This work was supported by the National Natural Science Foundation of China (Grant No. 12075143)

\end{document}